\documentclass[preprint2]{aastex61}
\newcommand{\source}{\src{}}
\newcommand{\src}{1E\,1048.1$-$5937}

\newcommand{\ps}{}

\shorttitle{Repeated outbursts of \source{}}
\shortauthors{Archibald et al.}

\usepackage{lineno, soul, color}
\usepackage{todonotes}

\begin{document}

\title{Two new outbursts and transient hard x-rays  from 1E~1048.1$-$5937}

\correspondingauthor{P. Scholz}
\email{paul.scholz@dunlap.utoronto.ca}

\author{R.F.~Archibald}
\affiliation{Department of Astronomy and Astrophysics, University of Toronto
	50 St. George Street, Toronto, ON M5S 3H4, Canada}  
\affiliation{Department of Physics \& McGill Space Institute, McGill University, 3600 University Street, Montreal QC, H3A 2T8, Canada}

\author{P.~Scholz}
\affiliation{Dunlap Institute of Astronomy and Astrophysics,
University of Toronto,
	50 St. George Street, Toronto, ON M5S 3H4, Canada
}

\author{V.~M.~Kaspi}
\affiliation{Department of Physics \& McGill Space Institute, McGill University, 3600 University Street, Montreal QC, H3A 2T8, Canada}

\author{S.~P.~Tendulkar}
\affiliation{Department of Physics \& McGill Space Institute, McGill University, 3600 University Street, Montreal QC, H3A 2T8, Canada}

\author{A. P. Beardmore}
\affiliation{Department of Physics and Astronomy, University of Leicester, University Road, Leicester LE1 7RH, UK}


\begin{abstract}
Since its discovery, \src{} has been one of the most active magnetars, both in terms of radiative outbursts, and changes to its spin properties. 
Here we report on a continuing monitoring campaign with  the {\it Neil Gehrels Swift Observatory} X-ray Telescope in which we observe two new outbursts from this source. 
The first outburst occurred in 2016 July, and the second in 2017 December, reaching peak 0.5-10\,keV absorbed fluxes of $3.2^{+0.2}_{-0.3}\times 10^{-11}$ erg\,s$^{-1}$\,cm$^{-2}$ and $2.2^{+0.2}_{-0.2}\times10^{-11}$\,erg\,s$^{-1}$\,cm$^{-2}$, respectively, factors of 
$\sim$5 and $\sim 4$ above the quiescent flux.
Both new outbursts were accompanied by spin-up glitches with amplitudes of  $\Delta\nu= 4.47(6)\times10^{-7}$\,Hz and  $\Delta\nu= 4.32(5)\times10^{-7}$\,Hz, respectively.
Following the 2016 July outburst, we observe, as for past outbursts, a period of delayed torque fluctuations, which reach a peak spin-down of $1.73\pm0.01$ times the quiescent rate, and which dominates the spin evolution compared to the spin-up glitches. 
We also report an observation near the  peak of the first of these outbursts  with {\it NuSTAR} in which \ps{hard X-ray emission is detected from the source.}
This emission is well characterized by an absorbed blackbody plus a broken power law, with a power-law index above $13.4\pm0.6$\,keV of $0.5_{-0.2}^{+0.3}$, similar to those observed in both persistent and transient magnetars.  The hard X-ray results are broadly consistent with models of electron/positron cooling in twisted magnetic field bundles in the outer magnetosphere.  However the repeated outbursts and associated torque fluctuations in this source remain puzzling.
\end{abstract}

\section{Introduction}
\src{}, 
one of the original ``anomalous X-ray pulsars'' \citep[AXPs;][]{1995ApJ...442L..17M}, 
is now classified as part of a small class of pulsars known as magnetars  -- neutron stars which display behavior thought to be powered by their immense magnetic fields.
For a recent review of magnetars see e.g.\,\cite{2017ARA&A..55..261K} or \cite{2018MNRAS.474..961C}.
A list of known magnetars is available at the {\it McGill Online Magnetar  Catalog} \citep{2014ApJS..212....6O}\footnote{\url{www.physics.mcgill.ca/\~pulsar/magnetar/main.html}}.

\src{} was discovered as a persistent X-ray source, with a pulse period of 6.4\,s, using the {\it Einstein X-ray Observatory} \citep{1986ApJ...305..814S}.
In the following decade, \src{} was occasionally observed with various X-ray missions and, by the mid-1990s, it was noticed that the spin-down rate was variable by  order unity \citep{1995ApJ...455..598M}.
\ps{X-ray flux variability in \src{} was first noted by \citet{2004ApJ...608..427M}.}
Starting in 1997,  \src{} was monitored regularly with the {\it Rossi X-ray Timing Explorer (RXTE)}, until the decommissioning of {\it RXTE} in 2012 \citep{2001ApJ...558..253K, 2004ApJ...609L..67G, 2014ApJ...784...37D}, and \ps{was monitored on a regular basis \citep{2015ApJ...800...33A} with the {\it Neil Gehrels Swift X-ray Telescope} (XRT) until
2018}. 

During this long-term monitoring, \src{} has been one of the most active known magnetars.
It has exhibited four long-term flux flares, as well as several magnetar-like bursts, and pulse profile changes.
Perhaps the most striking behavior in \src{} is the dramatically changing spin-down rate, which seems to occur regularly following its radiative outbursts \citep{2004ApJ...609L..67G, 2014ApJ...784...37D, 2015ApJ...800...33A}.
While many magnetars have been shown to have sudden timing changes associated with flux increases \citep[e.g.][]{2012ApJ...750L...6P, 2014ApJ...784...37D}, the repeated observation of an increased and variable torque  following each observed flux flare is as yet unexplained \citep{2015ApJ...800...33A}.
Counting the 2016 July outburst reported here, \src{} has now repeated this unusual behavior -- an X-ray outburst followed by delayed torque oscillations -- four times, each separated by $\sim$1700 days. 

Here we report on two X-ray outbursts and subsequent torque variations in \src.
The first of these in 2016 July occurred with a delay from the previous outburst consistent with that predicted by \cite{2015ApJ...800...33A}.
The second outburst, in 2017 December,  does not follow this timescale, however, as we show, it is less energetic than the major outbursts, and decays with a shorter timescale.

We also report the results of a new {\it NuSTAR} hard X-ray observation during the 2016 July outburst, wherein \src{} is \ps{detected above 20\,keV,} displaying the hard X-ray tail that is ubiquitous among the magnetar class. 

\section{Observations}

\subsection{{\it Swift} XRT Monitoring}
\label{sec:xrt}
\source{} \ps{was} monitored regularly with the {\it Swift}-XRT since  2011 July as part of a campaign to study several magnetars \citep[see e.g.][]{2014ApJ...783...99S,2015ApJ...800...33A, 2017ApJ...834..163A}.
The XRT was operated in Windowed-Timing (WT) mode for all observations,
having a time resolution of $1.76\;$ms, and only one dimension of spatial resolution.

Data were downloaded from the HEASARC \emph{Swift} archive, reduced using the  {\tt xrtpipeline} standard reduction script, and time-corrected to the Solar System Barycenter using {\tt HEASOFT v6.22}.
Following this, we processed the data in the same manner described by \cite{2017ApJ...834..163A}.

Observations, typically 1--1.5~ks long, were taken in groups of three, with the first two observations within approximately 8 hours of each other and the third approximately a day later.
This observation strategy was adopted due to the source's prior unstable timing behavior, in which maintaining phase coherence using a longer cadence was only possible for several-month intervals \citep{2001ApJ...558..253K, 2009ApJ...702..614D}.  
In total, 655 XRT observations totaling 1.0\,Ms of observing time spanning 2011 July through 2018 April were analyzed in this work.

\subsection{{\it NuSTAR} Observation}
Following the detection of the first new outburst reported in \S\ref{sec:longflux}, we received {\it NuSTAR} Director's Discretionary Time (DDT)  to observe \src{} in outburst.
The {\it NuSTAR} observation (obsid 90202032002) was taken on 2016 August 5 (MJD 57605) with an exposure time of 55\,ks.

{\it NuSTAR} data were reduced using the {\tt nupipeline} scripts, using {\tt HEASOFT v6.20} and time-corrected to the Solar System Barycenter.
Source events were extracted within a 1$'$ radius around the centroid.
Background regions were selected from the same detector as the source location, and spectra were extracted using the {\tt nuproducts} script.

Using {\tt grppha}, channels 0--35 ($<3$ keV) and 1935--4095 ($> 79$ keV) were ignored, and all good channels were binned to have a minimum of one count per energy bin.

\begin{figure}
	\center
	\includegraphics[width=\columnwidth]{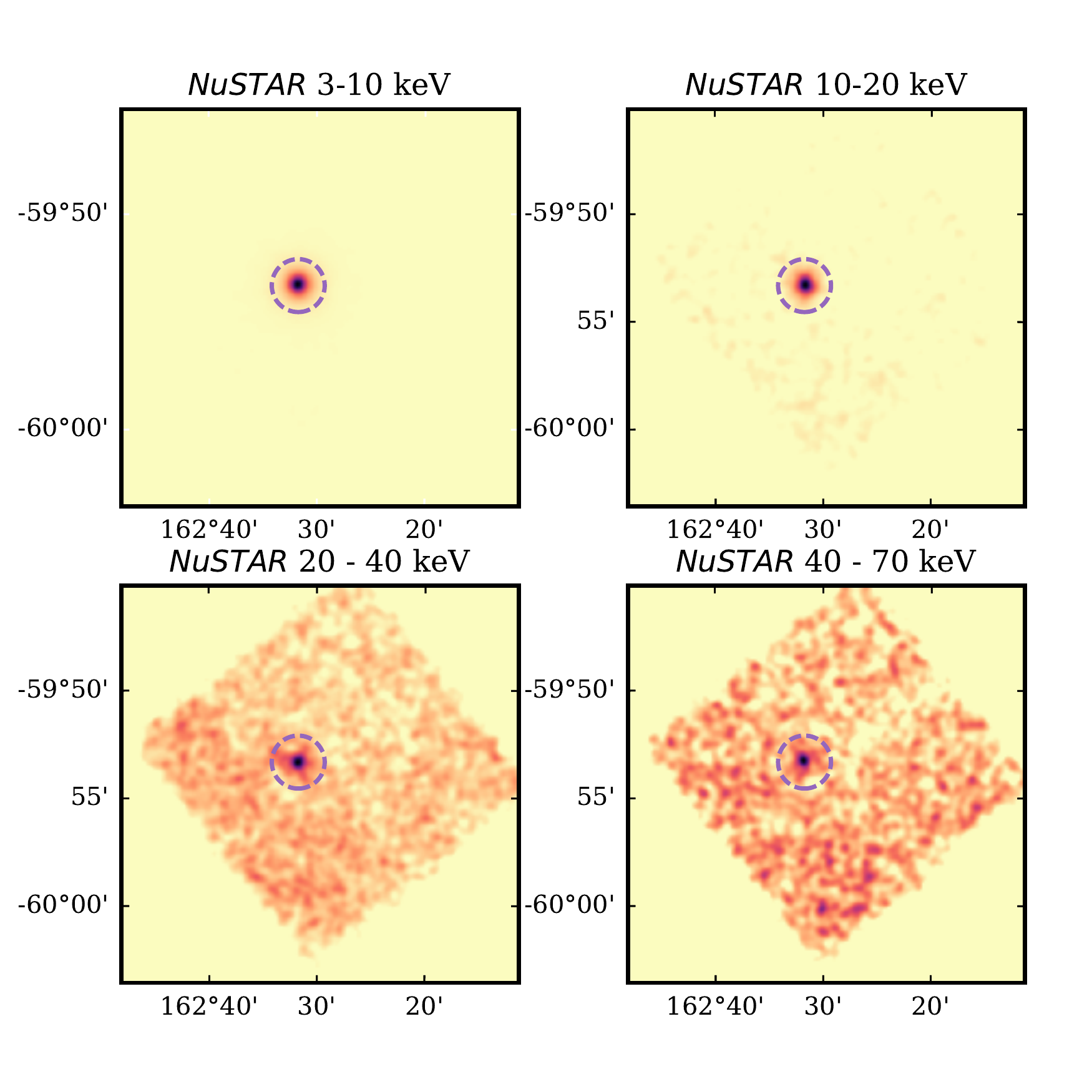}
	\caption{\ps{{\it NuSTAR} X-ray images in various energy bands of \source{} in outburst combining data from both focal plane modules.} The images have been smoothed with a Gaussian with a width of 4 pixels \ps{(10\arcsec)}. The position of \src{} is indicated by the dashed purple circle.}
	\label{fig:images}
\end{figure}

As shown in Figure~\ref{fig:images}, \src{} is clearly detected across the {\it NuSTAR} band, including at energies above 20\,keV allowing the spectral analysis described in \S\ref{sec:hard}.

\section{Flux \& Spectral Evolution}
\subsection{Long-term Evolution}
\label{sec:longflux}
Following the data reduction described in \S\ref{sec:xrt}, we fit the XRT observations using an absorbed blackbody model.
$N_\mathrm{H}$ was held constant at $5.8\times 10^{21}$\,cm$^{-2}$, the best-fit value for the source  before the 2012 outburst. 
Observations within one day of each other were grouped for this analysis.

Several individual observations, most notably in 2012 November, are significantly elevated from the long-term trend.
These are most likely due to catching \src{} during a period of post-burst tail emission lasting several kiloseconds, as reported by \cite{2014ApJ...790...60A}.

The long-term light-curve over the XRT campaign  is dominated by three outbursts.
We fit phenomenological models to the flux decay following each outburst, fixing the baseline flux to that measured before the 2011 December outburst, and fixing the outburst start time to that of the first observation \ps{with} elevated flux.
We first fit a single exponential decay, as well as power-law decays, to the flux following each outburst.
For the first two long outbursts, such single-component models did not adequately describe the data.

When fitting two-component models, those consisting of exponentials were statistically preferred to power-law models, using $\chi^2$ goodness of fit as a metric.
The optimal parameters  for a two-exponential model for each outburst are shown in Table~\ref{tab:outbursts}. 

\begin{figure}
	\centering
\includegraphics[width=0.9\columnwidth]{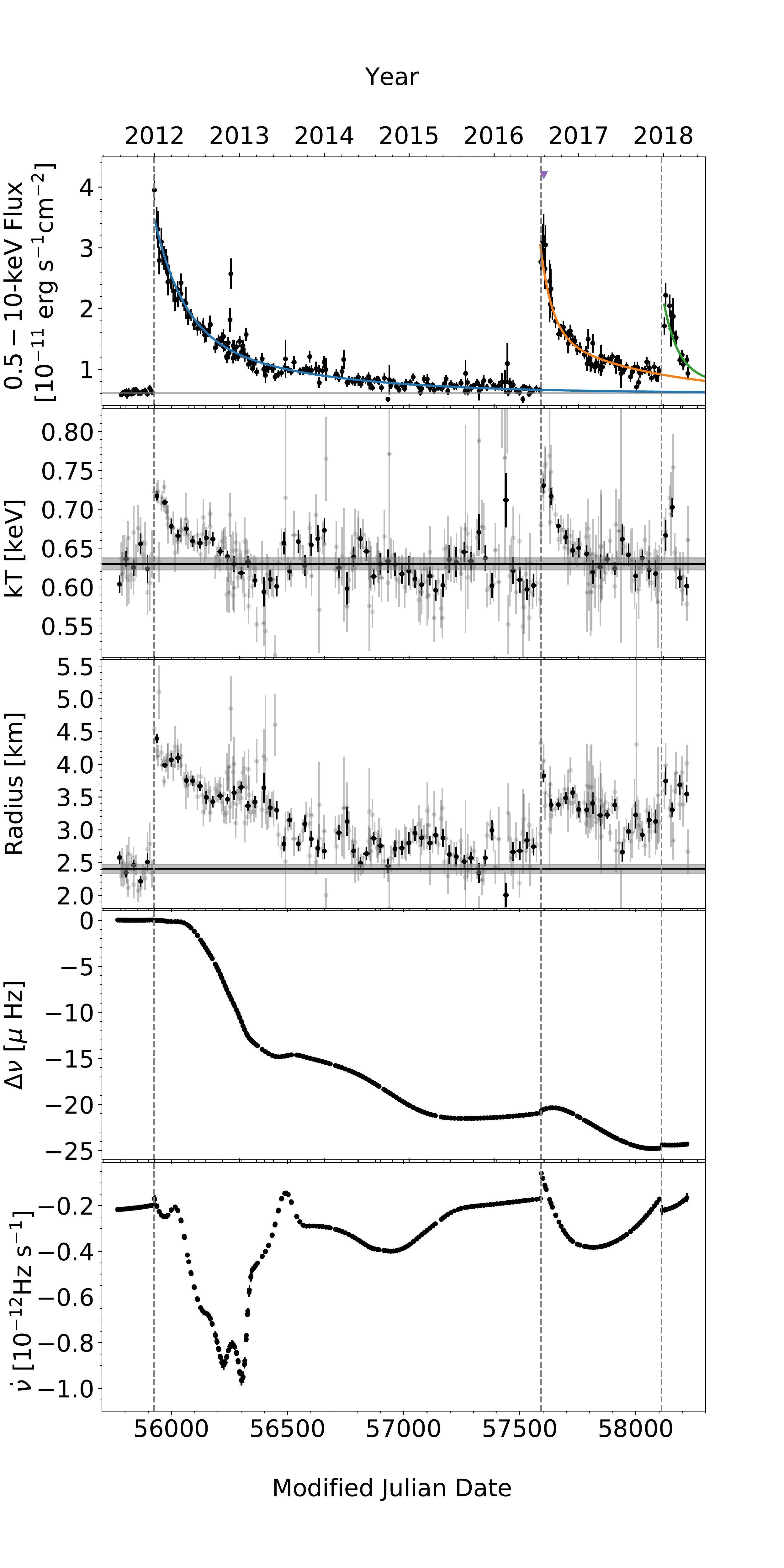}
	\caption{Flux and timing evolution of \src{} over the {\it Swift} campaign. The top panel shows the absorbed 0.5--10\,keV X-ray flux. The purple triangle indicates the time of the {\it NuSTAR} observation.  The best-fit models to the flux decays provided in Table~\ref{tab:outbursts} are shown as solid colored lines.
The second and third panels show the evolution of the blackbody spectral parameters, $kT$ and the radius, assuming a distance of 9\,kpc \citep{2006ApJ...650.1070D}. The light grey points show fits to observations grouped into days, and the black points combine observations into groups of one month.
  The bottom two panels show the evolution of the spin frequency, $\nu$, after the subtraction of the 2011 timing ephemeris, and that of the spin-down date, $\dot{\nu}$. The dashed vertical lines indicate the start of each outburst.}
\label{fig:swift_timing}	
\end{figure}

For the 2017 December outburst, as the 2016 July outburst had not yet fully decayed, we subtracted the best-fit model of that latter outburst before fitting.
As is evident from Figure~\ref{fig:swift_timing}, by the last observation reported here (2018 April), the effects of the 2017 December outburst have waned, with the last reported fluxes consistent with the extrapolation from the 2016 July outburst.

Note that for the two longest outbursts, both the short $\sim50$- and long $\sim500$-day exponential timescales  are consistent at the 1$\sigma$ level with each other.
In addition, the third outburst has a timescale consistent with the shorter $\sim50$-day timescale.
Also, within the limited available precision, the spectral variations are similar in the three outbursts (see Fig. 2).

\begin{table}
	\begin{center}
		\caption{Characterization of the flux decay during the 2016 and 2017 outbursts of \src{}.}
		\label{tab:outbursts}
		\begin{tabular}{c|c|c}\hline
$t_b$ 	& \ps{0.5--10\,keV} Flux Decay Fit$\star$  & $\chi^2_\nu$    \\ 
\hline
MJD 	& $10^{-11}$ erg\,s$^{-1}$\,cm$^{-2}$ &     \\ 
\hline
 55926	&$(1.1\pm0.15)$\,e$^{\frac{-(t-t_b)}{550\pm50}}$+$(1.9\pm0.13)$\,e$^{\frac{-(t-t_b)}{50\pm10}}$  & 1.1   \\ 
 57592	&$(1.0\pm0.13)$\,e$^{\frac{-(t-t_b)}{440\pm70}}$+$(1.5\pm0.16)$\,e$^{\frac{-(t-t_b)}{51\pm9}}$  &  0.75  \\ 
 58120	&$\dagger$ $(1.2\pm0.2)$\,e$^{\frac{-(t-t_b)}{62\pm12}}$ & 1.5  \\\hline

	\end{tabular} 
    $\star$ $t$ and $t_b$ are in units of days.
   
	$\dagger$ After subtraction of the flux decay fit from the July 2016 outburst; see \S\ref{sec:longflux}.
\end{center} 
\end{table}

\subsection{Hard X-rays in Outburst}
\label{sec:hard}

\begin{table}
	\begin{center}
		\caption{{\it NuSTAR} \& {\it Swift}-XRT spectrum of \src{} in outburst.}
		\label{tab:spec}
		\begin{tabular}{lc}
			\multicolumn{2}{c}{Absorbed Blackbody \& Broken Power-law} \\
			\hline
			Parameter &  Value \\\hline
			$N_\mathrm{H}$ ($10^{22}\,\mathrm{cm^{-2}}$) & $3.7\pm0.3$\\
$C_{NuSTAR}\tablenotemark{a} $ & $0.85\pm0.05$\\
			$kT_\mathrm{BB}$ (keV) & $0.88\pm0.02$ \\
			$\Gamma_S$ & $4.4\pm0.1$  \\
			$\Gamma_H$ & $0.5_{-0.2}^{+0.3}$  \\
			Break Energy (keV) & $13.4^{+0.6}_{-0.6}$ \\
			C-Stat/$\mathrm{dof}$ & 1809.7/1965 \\
            Goodness\tablenotemark{b} & $49.7\%$ \\
			Flux (0.5--10\,keV)\tablenotemark{c} & $31.2^{+0.7}_{-1.5}$ \\
			Flux (3--79\,keV)\tablenotemark{c} & $20.^{+1}_{-2}$  \\ 
 
			Flux (20--79\,keV)\tablenotemark{c} & $4.8^{+1.4}_{-1.2}$  \\\hline
			
		\end{tabular}
        \tablenotetext{\rm a}{Fitted relative normalization for {\it NuSTAR}.}
        \tablenotetext{\rm b}{Percentage of C-Stat statistic simulation trials from model parameters that are less than the fit statistic.}
		\tablenotetext{\rm c}{Absorbed flux in units of $10^{-12}\,\mathrm{erg\,cm^{-2}\,s^{-1}}$.}
	\end{center}
\end{table} 

A {\it NuSTAR} observation was taken approximately 13 days after the {\it Swift}-XRT-detected flux increase, as indicated in Figure~\ref{fig:swift_timing}.
We first verified that there were no short, magnetar-like bursts contaminating the data by conducting a burst search following the method described by \cite{2011ApJ...739...94S}.
To constrain the soft X-ray spectrum, we co-fit the {\it NuSTAR} observation with the {\it Swift} observations (observation ids 00032923252 \& 254) taken  on August 5--8 2016, coincident to within days of the epoch of the {\it NuSTAR} observation.

We used Cash statistics \citep{1979ApJ...228..939C} for fitting and parameter estimation of the unbinned data. 
$N_\mathrm{H}$ was fit using the {\tt tbabs} model with \texttt{wilm} abundances \citep{2000ApJ...542..914W} and \texttt{vern} photoelectric cross-sections \citep{1996ApJ...465..487V}. 

\begin{figure}
	\center
	\includegraphics[width=\columnwidth]{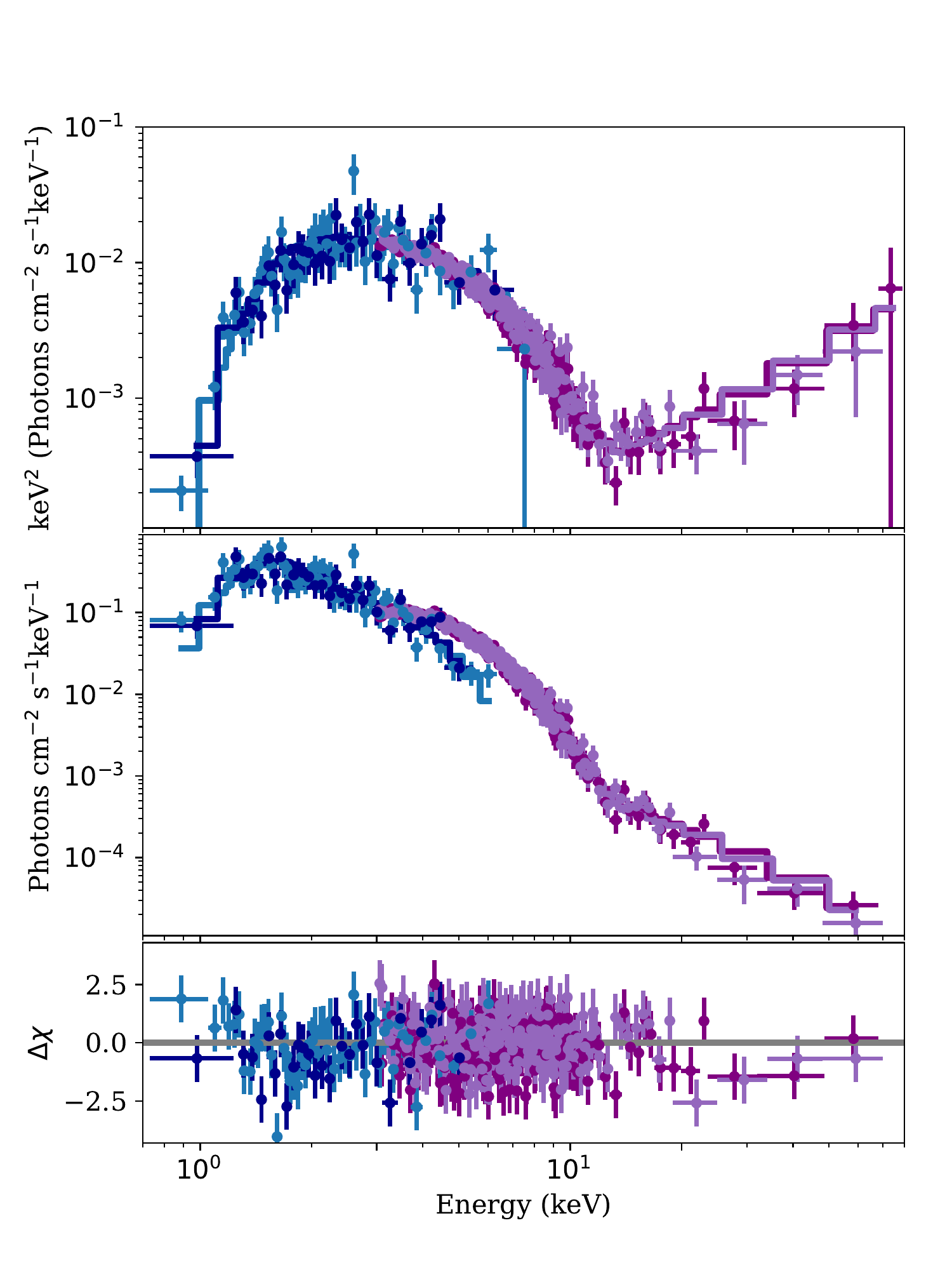}
	\caption{X-ray spectra of \source{} in outburst. In all panels, the blue data points are from {\it Swift}-XRT, and purple data points from {\it NuSTAR}. The top panel shows the spectral energy distribution, the middle panel shows the observed spectrum
and the bottom panel displays the residuals of the data relative to the model presented in Table~\ref{tab:spec}. }
	\label{fig:spec}
\end{figure}

\ps{\src{} is detected above 20\,keV} with a background subtracted 20--79-keV count rate of $(5.3\pm0.6)\times10^{-3}$ photons per second. 
The spectrum is well fit by an absorbed blackbody and broken power law; the best-fit parameters are shown in Table~\ref{tab:spec}.
Here, $\Gamma_S$ and $\Gamma_H$ refer to the power-law index below and above the break energy, respectively.
The X-ray spectrum and residuals are shown in Figure~\ref{fig:spec}.
All the uncertainties in the spectral parameters are quoted at 90\% confidence.

In a 2013 {\it NuSTAR} observation of \src{} in relative quiescence, neither \cite{2015ApJ...815...15W} nor \cite{2016ApJ...831...80Y} found any evidence of X-ray flux from \src{} above 20\,keV, setting a 3\,$\sigma$ upper limit on the total, phase-averaged flux in the  20--79\,keV band of $\sim$3--4$\times 10^{-12}\,\mathrm{erg\,cm^{-2}\,s^{-1}}$, just below our detection of a flux of $4.8^{+1.4}_{-1.2} \times 10^{-12}\,\mathrm{erg\,cm^{-2}\,s^{-1}}$. 

\subsection{{\it NuSTAR} Pulsed Flux}

\begin{figure}
	\center
	\includegraphics[width=0.9\columnwidth]{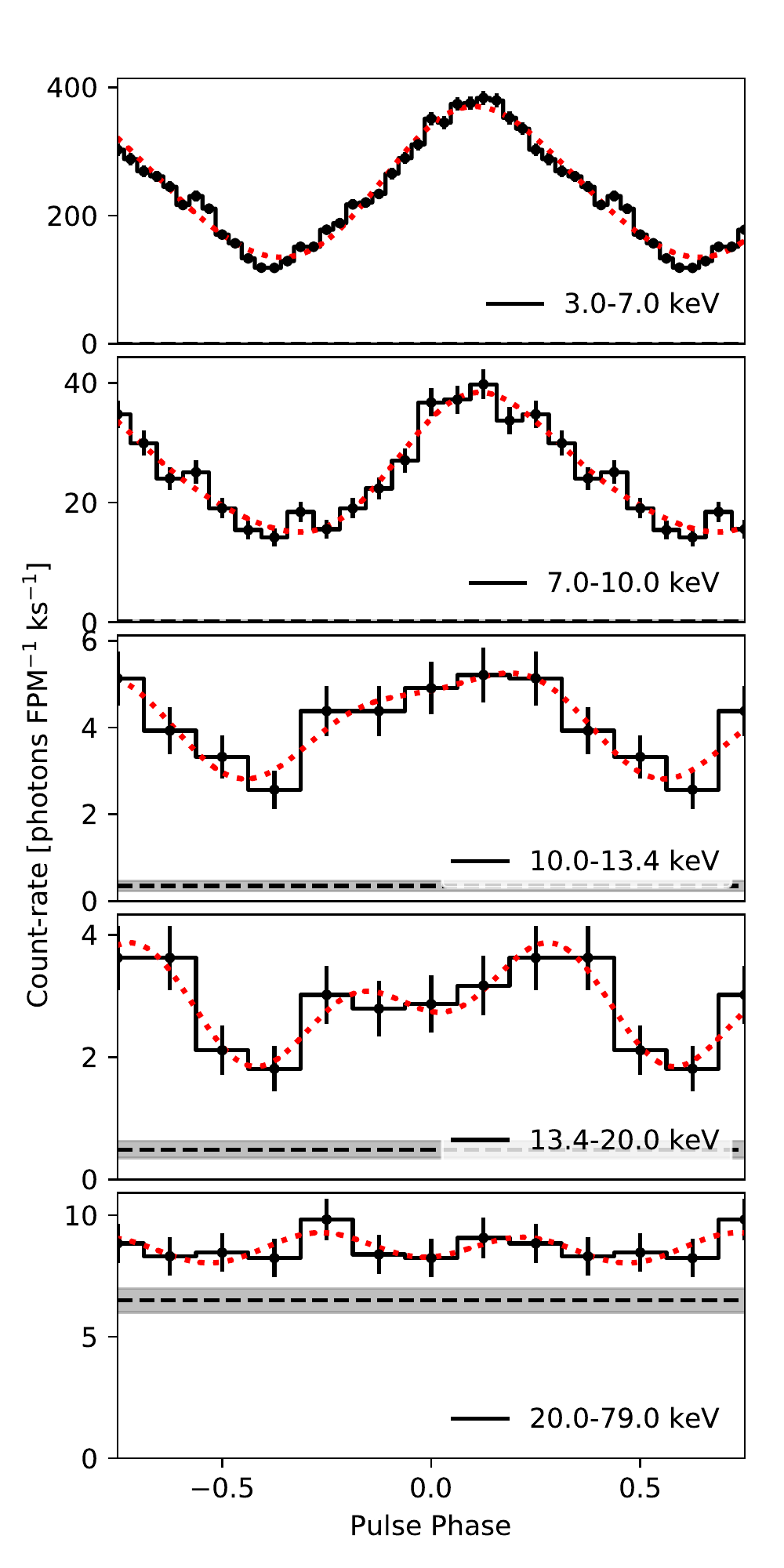}
	\caption{Pulse profile of \source{} in the {\it NuSTAR} observation in various energy bands. In all panels, the black dashed line represents the background count rate, and the red dashed line shows the H-test preferred pulse profile.}
	\label{fig:pulse}
\end{figure}

In Figure~\ref{fig:pulse}, we show the pulse profiles of \src{} during the {\it NuSTAR} observation in units of photons per kilosecond per Focal Plane Module (FPM), folded using the timing solution from the {\it Swift} campaign (see \S\ref{sec:timing}).
We calculated the RMS pulsed fraction of \src{} in several energy bands, using the method described in the appendix of \cite{2015ApJ...807...93A}.
To determine the significance of the pulsed signal, we used the H-test \citep{1989A&A...221..180D}.
Motivated by the spectral break in the power law at  $13.4^{+0.6}_{-0.6}$\,keV, we used this value as a fiducial cut to search for a pulsed signal in the hard X-ray band.
A pulsed signal is detected up to 20\,keV, and no significant pulsations are seen above this energy. 
In Table~\ref{tab:pulse} we report the H-test false-alarm-probabilities (P$_{FA}$), and pulse fractions, where upper limits are given at the 99\% confidence level.
Due to a paucity of pulsed counts in the hard X-ray band, we can neither comment on the energy-dependence of the pulsed fraction, nor do meaningful phase-resolved spectroscopy of \src{}.

\begin{table}
	\begin{center}
		\caption{{\it NuSTAR} Pulsed Flux from \src{}.}
		\label{tab:pulse}
		\begin{tabular}{c|c|c}\hline
Energy Range &H-test P$_{FA}$	&RMS Pulsed Fraction$\dagger$   \\ 
\hline
keV 	&  & \%   \\ 
\hline
3--7 	&0& $51\pm1$    \\ 
7--10 	&$1\times10^{-47}$& $48\pm3$     \\ 
10--13.4 	&$1\times10^{-3}$& $28\pm8$    \\ 
13.4--20 	&0.03& $30\pm10$     \\ 
20--79 	&0.9&  $<80$   \\ \hline

	\end{tabular} 
   
	$\dagger$ After background subtraction.
\end{center} 
\end{table}

\section{Timing Analysis}
\label{sec:timing}
The processed individual XRT photons were used to derive a pulse time-of-arrival (TOA) for each observation. 
The rotational phase ($\phi_i$) of every photon in the observation was calculated, assuming the best prior timing model.
The TOAs were created using a Maximum Likelihood (ML) method as described by  \cite{2009LivingstoneTiming} and \cite{2012ApJ...761...66S}.

These TOAs were fitted to a timing model in which the phase $\phi$ as a function of time $t$ is described by a Taylor expansion:
\begin{equation}
\phi(t) = \phi_0+\nu_0(t-t_0)+\frac{1}{2}\dot{\nu_0}(t-t_0)^2+\frac{1}{6}\ddot{\nu_0}(t-t_0)^3+\cdots
\end{equation}
where $\nu$ is the rotational frequency of the pulsar.
This was done using the {\tt tempo2} pulsar timing software package \citep{2006MNRAS.369..655H}. 

As the frequency derivative of \src{} changes by up to an order of magnitude on $\sim$months time scales, we first created overlapping timing solutions with {\tt tempo2} to determine a relative pulse number for each TOA.  Then, using the overlapping regions to ensure the same number of rotations in each solution, these solutions were merged, allowing the establishment of absolute pulse numbers throughout the entire {\it Swift} campaign.
 
In order to determine the local timing behavior of \src{}, we fit splines to these absolute pulse numbers \citep[see][]{splineref}, using a method similar to that described by \cite{2014ApJ...784...37D}, using piecewise polynomials of degree $n=3$ weighted by the inverse square error on the pulse phase.
To determine uncertainties, we refit these splines 1000 times after adding Gaussian noise to the pulse numbers, using their measured pulse phase uncertainties.
The resulting spin frequencies and frequency derivatives are shown in Figure~\ref{fig:swift_timing}.
The plotted error bars, typically comparable to the size of the points, indicate the 68\% confidence regions.

We detected a spin-up glitch
coincident with the 2016 July flux increase.
As is evident in Figure~\ref{fig:swift_timing}, the timing parameters of \src{} are not stable.
To measure the size of the glitch, we fit a simple timing solution in the interval MJD 57400--57668, consisting of $\nu$ and $\dot{\nu}$ as well as a glitch in $\nu$ with the epoch fixed to that of the flux increase.
This yields a glitch with $\Delta\nu= 4.47(6)\times10^{-7}$\,Hz ($\Delta\nu/\nu= 2.89(4)\times10^{-6}$).
The above epoch bounds were chosen to have a reduced $\chi^2\sim1$ and to result in no visible trends in the residuals.
We note that the actual timing evolution is more complicated, as is evident in Figure~\ref{fig:swift_timing}.

In the same manner, we also find a glitch coincident with the 2017 December flux increase.
Fitting a simple timing solution in the interval MJD 58000--58200 with the epoch fixed to that of the flux increase gives a glitch having $\Delta\nu= 4.32(5)\times10^{-7}$\,Hz ($\Delta\nu/\nu= 2.79(3)\times10^{-6}$).
Again, note that the actual timing evolution is more complicated (Fig.~\ref{fig:swift_timing}).

The influence of these glitches on the long-term spin down of the pulsar is far smaller than the integrated effect of the varying torque.
Collectively, the two glitches change $\nu$ by $\sim8.8\times10^{-7}$\,Hz while the added spin-down variations have contributed $\sim-2\times10^{-5}$\,Hz.

\section{Discussion}
\subsection{Hard X-ray Component}

\ps{Here we have presented the detection of \src{} at energies above 20\, keV. This, however, is not the first 
high energy detection of the source. \citet{2008A&A...477L..29L} detected \src{} at 22--100\,keV with {\it INTEGRAL} 
during observations of $\eta$ Carinae. Their observation totals 1.1\,Ms and is drawn from several observing epochs,
but one of those epochs (MJD 52787--52827) corresponds to the peak of the 2001--2002 outburst of \src{}. This is therefore
consistent with the picture of \src{} being bright in hard X-rays during outburst.
}

Hard X-ray emission from magnetars is ubiquitous in persistently bright magnetars \citep[e.g.][]{2006ApJ...645..556K,2014ApJ...789...75V, 2017ApJ...851...17Y, 2017ApJS..231....8E}.
Additionally, in transient magnetars, similar  hard-X-ray components are observed near epochs of enhanced flux.
 For example, in SGR\,0501$+$4516, for which, in the first four days of an outburst, {\it Suzaku} detected a hard power law with $\Gamma=0.79_{-0.18}^{+0.20}$ \citep{2010ApJ...715..665E} -- similar to the spectrum we have observed in \src{}.
As well, in SGR\,1935+2154 \citep{2017ApJ...847...85Y}, a hard X-ray component was observed at the peak flux of an outburst.  Thus the phenomenon of a transient hard X-ray component appearing in outburst seems common for the magnetar class.

This hard X-ray emission is thought to be due to  decelerating electron/positron flow in large twisted magnetic loops of the pulsar magnetosphere \citep{2013ApJ...762...13B}. 
In this picture, the flux evolution of magnetars following outbursts involves the untwisting of the magnetosphere \citep[e.g][]{2009ApJ...703.1044B, 2013ApJ...774...92P, 2017ApJ...844..133C}. 
The transient hard-X-ray emission we observed in \src{}, and other magnetars in outburst, is then consistent with this picture where hard-X-ray emission is only detectable during the peak of this outburst when the magnetosphere is maximally twisted. 
We would then generally expect the evolution of the hard X-ray flux to proceed on a similar timescale to that of the soft X-ray flux \citep{2017ApJ...844..133C}.
Future systematic hard X-ray observations of magnetars in outburst are needed to put this to the test,
although the hard X-ray relaxation of the high-magnetic-field radio pulsar PSR~J1119$-$6127 has recently been shown to proceed on a time scale similar to that of the soft X-ray relaxation post-outburst \citep{2018ApJ...869..180A}.


A correlation has been observed between the surface magnetic field (or alternately the spin-down rate) of a magnetar and its hard X-ray power-law index \citep{2010ApJ...710L.115K, 2017ApJS..231....8E}.
Indeed, \cite{2010ApJ...710L.115K} predicted that $\Gamma_H$ for \src{} should fall between 0--1, albeit in quiescence.  Interestingly, this is in agreement with our measurement of $\Gamma_H = 0.5^{+0.3}_{-0.2}$ in outburst.

In \cite{2017ApJS..231....8E}, the hardness ratio of fluxes in the 15--60\,keV and 1--10\,keV bands is shown to be correlated with the spin-down rate of the magnetar.
If we take the quiescent spin-down rate of \src{} ($\sim9\times10^{-12}$\,s\,s$^{-1}$), the predicted hardness ratio for \src{} is $\sim0.4$.
We measure $F_{15-60\,\mathrm{keV}} = (3.2\pm0.6)\times10^{-12}$\,erg\,cm$^{-2}$\,s$^{-1}$ and $F_{1-10\,\mathrm{keV}} = (31\pm 1)\times10^{-12}$\,erg\,cm$^{-2}$\,s$^{-1}$ for a hardness ratio of $0.10\pm 0.02$ which is broadly consistent with the trend, especially given the large fluctuations in $\dot{P}$ observed in \src{}, as well as the scatter in the observed distribution \citep{2017ApJS..231....8E}.

\begin{figure*}
	\centering
	\includegraphics[width=1.9\columnwidth]{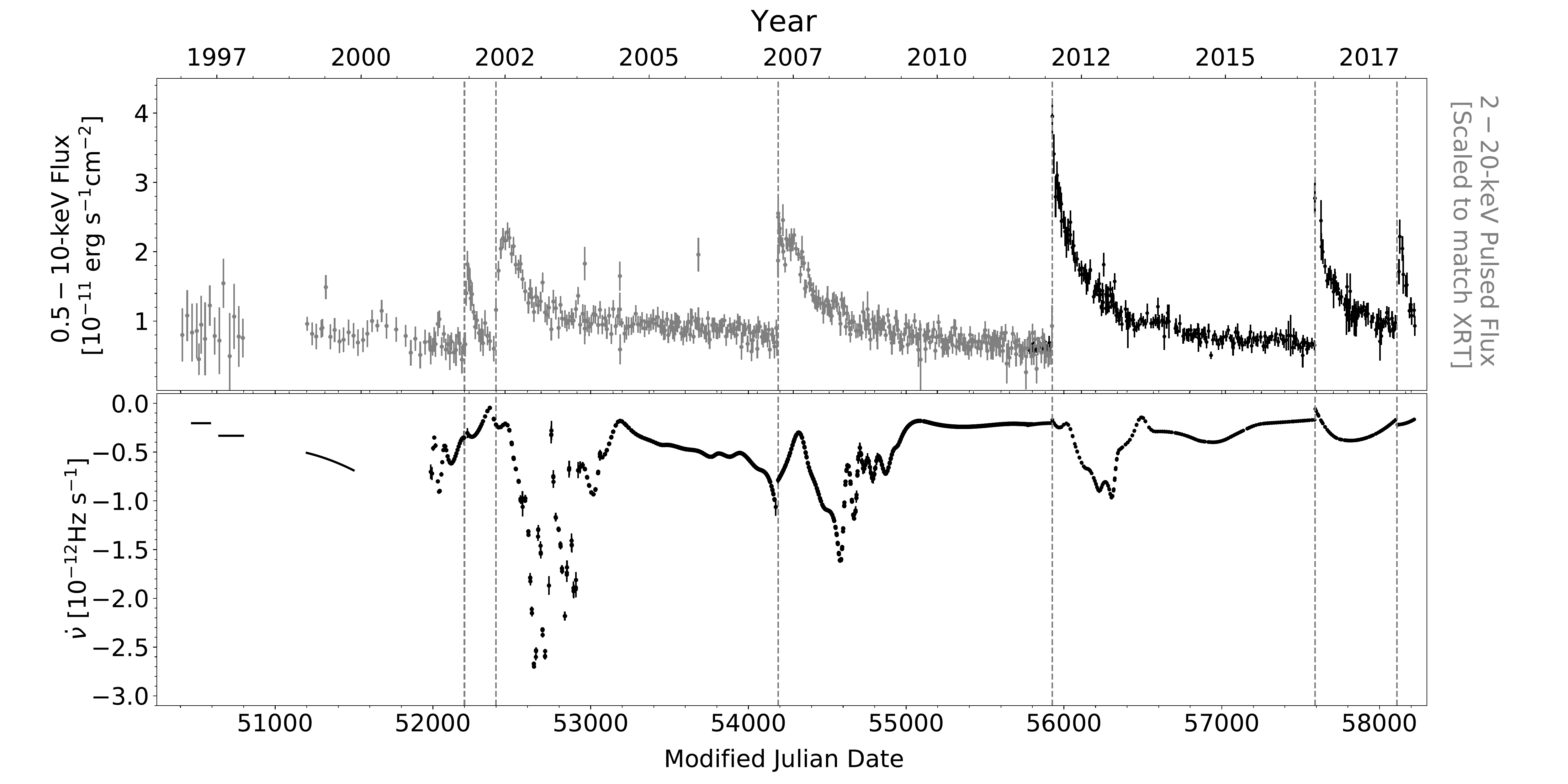}
	\caption{Flux and timing evolution of \src{} over the combined {\it RXTE} and {\it Swift} campaigns. The top panel shows the absorbed 0.5--10\,keV X-ray flux in black, and the {\it RXTE} 2--20 pulsed flux measured with the proportional counter array \citep{1996SPIE.2808...59J} in grey, scaled to match the {\it Swift} total flux. The bottom panel shows the evolution of the spin-down date, $\dot{\nu}$. In both panels, the dashed vertical lines indicate the start of a flux increase. The {\it RXTE} data are from \cite{2014ApJ...784...37D}, with the timing solutions pre-2000 from \cite{2001ApJ...558..253K}.}
	\label{fig:long_timing}
\end{figure*}

\subsection{Repeated outbursts \& torque changes in \src{}}
In Figure~\ref{fig:long_timing} we show the last 20 yr of evolution in the X-ray flux, and spin-down rate for \src{}, as monitored by {\it RXTE} \footnote{ The {\it RXTE} data presented here are reproduced from \cite{2014ApJ...784...37D}, with the timing solutions pre-2000 from \cite{2001ApJ...558..253K}.} and {\it Swift}.
Note that the fluxes are in different energy bands (0.5--10\,keV vs 2--20\,keV), and that {\it RXTE} fluxes are pulsed only, and have been scaled to match the {\it Swift} flux during the period of overlap.

The time delay between the 2011 December and the 2016 July outbursts was $1670\pm10$ days.  This can be compared to separations of $1800\pm10$ and $1740\pm10$ days between the prior flares as discussed by \citet{2014ApJ...784...37D} and \citet{2015ApJ...800...33A}.
While this outburst timing is consistent with the quasi-periodicity suggested in \cite{2015ApJ...800...33A}, the occurrence of the 2017 December outburst suggests that this repeated time scale is spurious.  However, this last outburst is decaying on a faster timescale than the major outbursts on which the claimed quasi-periodicity is based -- similar to the precursor flare noted in 2001 \citep[e.g.][]{2008ApJ...677..503T, 2014ApJ...784...37D}.
It will be interesting to continue monitoring \src{} to see if there is another outburst on the timescale the quasi-periodicity predicts, i.e. in $\sim$2021.

Additionally, the torque variations following the 2016 July outburst follow the trend of decreasing amplitude noted in \cite{2015ApJ...800...33A}.
Following the four major outbursts observed thus far, the peak torque reached values of 12.3(1), 7.32(5), 4.4(1), and finally 1.73(1) times higher than the quiescent rate. 
The monotonic decrease in amplitude of these unexplained torque variations is curious, as it implies that our monitoring of \src{} was started at a special time, perhaps after a major but unobserved event.  If the decline continues, by the next outburst, the torque variations should be smaller than order unity times the quiescent value.  However, the monotonic decrease may also be purely coincidental.  Further monitoring will be  illuminating.

\ps{While the repetition, and monotonic decline in amplitude, of the torque variations} from \src{} are striking and unique,   rapid, extreme variability in the torque ($\dot{\nu}$) evolution appears to be a common feature following magnetar outbursts.
In addition to that observed now repeatedly in \src{}, similar variations have been observed in
1E\,1547$-$5408 \citep{2012ApJ...748....3D},
PSR\,J1622$-$4950 \citep{2017ApJ...841..126S, 2018ApJ...856..180C}, and in
XTE\,1810$-$197 \citep{2016ApJ...820..110C}. 
Thus, in a large fraction of magnetar outbursts for which the spin-down rate has been tracked for over a decade, these extreme torque variations are observed, and can dominate the long-term spin evolution of these sources.

In the magnetar model, increased torque associated with outbursts, just as the enhanced hard X-ray emission, is due to a twist in the magnetosphere \citep[e.g][]{2002ApJ...574..332T, 2009ApJ...703.1044B}.
As the spin-down rate of the star is dominated by the relatively small number of open field lines, there is no reason for a strict correlation  between the hard X-ray emission and spin-down rate, as it depends on the geometry of the magnetosphere \citep{2009ApJ...703.1044B, 2017ARA&A..55..261K}.
In the untwisting model, the spin-down rate of the star is only affected once the twist reaches an amplitude of $\sim 1$\,radian. 
The delay between the peak X-ray flux and peak torque of $\sim100$\,days observed in \src{} would then be due to the initial twist not exceeding this threshold
value \citep{2009ApJ...703.1044B}.

\section{Conclusions}
We have presented long-term X-ray observations of \src{} during which we observe two new outbursts of this source in 2016 July and 2017 December.
Associated with these outbursts, we find spin-up glitches having $\Delta\nu/\nu$ of order $ 10^{-6}$, although the long-term spin evolution is dominated by a strongly fluctuating spin-down rate.
We also report a transient hard X-ray component of \src{} observed with {\it NuSTAR} near the peak of the 2016 July outburst, with emission up to $\sim70$\,keV, and pulsed emission observed up to 20~keV.  The spectrum and pulse properties of this hard emission are qualitatively consistent with emission models involving cooling of electron/positron pairs in large, twisted magnetic loops in the outer regions of the stellar magnetosphere \citep{2013ApJ...762...13B}.  The repeating outbursts and associated large, delayed torque variations, and their possible monotonic decline in amplitude in \src{} remain, however, puzzling.

\acknowledgements
R.F.A. acknowledges support from an  NSERC  Postdoctoral Fellowship.
P.S. is a Dunlap Fellow and an NSERC Postdoctoral Fellow. The Dunlap Institute is funded through an endowment established by the David Dunlap family and the University of Toronto. 
V.M.K. receives support from an NSERC Discovery Grant and Herzberg Award, the Centre de Recherche en Astrophysique du Qu\'ebec, an R. Howard Webster Foundation Fellowship from the Canadian Institute for Advanced Study, the Canada Research Chairs Program and the Lorne Trottier Chair in Astrophysics and Cosmology.
A.P.B. acknowledges funding from the UK Space Agency.
The authors thank the operations team of {\it NuSTAR} for approving a rapid turn-around DDT.
We thank the {\it Swift} team for approving our ToO requests to monitor \src{}, and  other magnetars over the years.
This research has made use of data obtained through the High Energy Astrophysics Science Archive Research Center Online Service, provided by the NASA/Goddard Space Flight Center.
This work made use of data from the {\it NuSTAR} mission, a project led by the California Institute of Technology, managed by the Jet Propulsion Laboratory, and funded by the National Aeronautics and Space Administration.


\facilities{{\it NuSTAR, Swift}}
	
\software{\newline {\tt numpy} \citep{2011CSE....13b..22V},
          \newline {\tt astropy} \citep{2013A&A...558A..33A,2018AJ....156..123A},
          \newline {\tt xspec} \citep{1996ASPC..101...17A},
          \newline {\tt heasoft} \citep{2014ascl.soft08004N}}
\bibliography{rfa}{}
\bibliographystyle{yahapj}

\end{document}